\documentclass{webofc}
\usepackage[varg]{txfonts}
\usepackage{hyperref}
\usepackage{url}
\hypersetup{colorlinks=true,citecolor=blue,urlcolor=blue,linkcolor=blue}
\usepackage{booktabs}
\usepackage[nolist,nohyperlinks]{acronym}
\usepackage[mathlines]{lineno}
\acrodef{HL-LHC}{High Luminosity Large Hadron Collider}
\acrodef{HEP}{High Energy Physics}
\acrodef{AGC}{Analysis Grand Challenge}
\acrodef{AF}{Analysis Facility}
\acrodefplural{AF}{Analysis Facilities}
\acrodef{ML}{Machine Learning}
\acrodef{WLCG}{Worldwide LHC Computing Grid}
\begin{document}
\title{The 200 Gbps Challenge: Imagining HL-LHC analysis facilities}
\author{
    \firstname{Alexander} \lastname{Held}\inst{1}\fnsep\thanks{\email{alexander.held@cern.ch}} \and
    \firstname{Sam} \lastname{Albin}\inst{2} \and
    \firstname{Garhan} \lastname{Attebury}\inst{2} \and
    \firstname{Kenneth} \lastname{Bloom}\inst{2} \and
    \firstname{Brian} \lastname{Bockelman}\inst{3} \and
    \firstname{Lincoln} \lastname{Bryant}\inst{4} \and
    \firstname{Kyungeon} \lastname{Choi}\inst{5} \and
    \firstname{Kyle} \lastname{Cranmer}\inst{1} \and
    \firstname{Peter} \lastname{Elmer}\inst{6} \and
    \firstname{Matthew} \lastname{Feickert}\inst{1} \and
    \firstname{Rob} \lastname{Gardner}\inst{4} \and
    \firstname{Lindsey} \lastname{Gray}\inst{7} \and
    \firstname{Fengping} \lastname{Hu}\inst{4} \and
    \firstname{David} \lastname{Lange}\inst{6} \and
    \firstname{Carl} \lastname{Lundstedt}\inst{2} \and
    \firstname{Peter} \lastname{Onyisi}\inst{5} \and
    \firstname{Jim} \lastname{Pivarski}\inst{6} \and
    \firstname{Oksana} \lastname{Shadura}\inst{2} \and
    \firstname{Nick} \lastname{Smith}\inst{7} \and
    \firstname{John} \lastname{Thiltges}\inst{2} \and
    \firstname{Ben} \lastname{Tovar}\inst{8} \and
    \firstname{Ilija} \lastname{Vukotic}\inst{4} \and
    \firstname{Gordon} \lastname{Watts}\inst{9} \and
    \firstname{Derek} \lastname{Weitzel}\inst{2} \and
    \firstname{Andrew} \lastname{Wightman}\inst{2}
}
\institute{
    University of Wisconsin--Madison, United States \and
    University of Nebraska--Lincoln, United States \and
    Morgridge Institute for Research, United States \and
    University of Chicago, United States \and
    University of Texas at Austin, United States \and
    Princeton University, United States \and
    Fermilab, United States \and
    University of Notre Dame, United States \and
    University of Washington, United States
}
\abstract{%
The IRIS-HEP software institute, as a contributor to the broader HEP Python ecosystem, is developing scalable analysis infrastructure and software tools to address the upcoming HL-LHC computing challenges with new approaches and paradigms, driven by our vision of what HL-LHC analysis will require. The institute uses a ``Grand Challenge'' format, constructing a series of increasingly large, complex, and realistic exercises to show the vision of HL-LHC analysis. Recently, the focus has been demonstrating the IRIS-HEP analysis infrastructure at scale and evaluating technology readiness for production.

As a part of the Analysis Grand Challenge activities, the institute executed a ``200 Gbps Challenge'', aiming to show sustained data rates into the event processing of multiple analysis pipelines. The challenge integrated teams internal and external to the institute, including operations and facilities, analysis software tools, innovative data delivery and management services, and scalable analysis infrastructure. The challenge showcases the prototypes --- including software, services, and facilities --- built to process around 200 TB of data in both the CMS NanoAOD and ATLAS PHYSLITE data formats with test pipelines.

The teams were able to sustain the 200 Gbps target across multiple pipelines. The pipelines focusing on event rate were able to process at over 30 MHz. These target rates are demanding; the activity revealed considerations for future testing at this scale and changes necessary for physicists to work at this scale in the future. The 200 Gbps Challenge has established a baseline on today's facilities, setting the stage for the next exercise at twice the scale.
}
\maketitle
\acresetall
\section{Introduction}

The upcoming \ac{HL-LHC} will deliver a dataset of collisions with a total size significantly exceeding that resulting from the Run-2 and ongoing Run-3 operation of the LHC.
Both the CMS~\cite{Software:2815292} and ATLAS~\cite{CERN-LHCC-2022-005} collaborations have studied the computing challenges associated with handling this data and outlined the need for R\&D to meet the computational requirements of physics analyses with \ac{HL-LHC} data.

The focus of the work reported here lies in the last stages of a \ac{HEP} analysis pipeline, in the realm of \textit{end user physics analysis}.
Earlier stages in the pipeline are commonly centrally organized and the end user part begins with centrally provided input files.
The subsequent workflow can vary a lot between analyses, but typically has to be run frequently from the design stage of a physics analysis all the way through to the publication of the results.
The turnaround time for this end user workflow increase with dataset size.
This poses the risk of large increases in the time it takes to perform a physics analysis at the \ac{HL-LHC}.
Mitigating this risk is one of the missions of the IRIS-HEP software institute.

\subsection{IRIS-HEP and the ``Grand Challenge'' format}

IRIS-HEP~\cite{irishep} is the Institute for Research and Innovation in Software for High Energy Physics.
Since 2018, IRIS-HEP has performed computing and software R\&D for the \ac{HL-LHC}: it targets a \textit{software upgrade} accompanying the detector hardware upgrades that are being pursued for the CMS and ATLAS experiments.
This work is being done in close collaboration with the experiments and computing facilities.
IRIS-HEP members are a mix of physicists, computer scientists, and engineers, distributed across many institutes in the United States.
They work on a broad range of topics, many of which are connected via ``Grand Challenges''.
These challenges outline a series of exercises of increasing complexity towards realistic \ac{HL-LHC} physics analysis scale.
A recent focus has been end user physics analysis: demonstrating IRIS-HEP analysis infrastructure at scale and evaluating technology readiness.
Two recent key projects are situated in this field:
\begin{itemize}
    \item \textbf{\ac{AGC}}~\cite{Held:2022sfw,acat_proceedings,Held:2024gwj,Kauffman:2024bov}: an end-to-end analysis pipeline that serves as an integration exercise,
    \item \textbf{200 Gbps Challenge}: demonstrating data throughput of 200 gigabit per second for physics analysis applications.
\end{itemize}

\subsection{Overall vision}

A central goal for IRIS-HEP is to empower physicists by minimizing the time-to-insight and thereby maximizing the \ac{HL-LHC} physics reach.
The time-to-insight is the turnaround time of physics analysis and incorporates time spent debugging, bookkeeping, and waiting for computing to finish.
The institute aims to tighten feedback and support cycles in particular among three groups of people:
\begin{itemize}
    \item physicists,
    \item software developers,
    \item analysis facility experts.
\end{itemize}
An efficient and scalable implementation of a physics analysis pipeline requires expertise in these three areas.
By having people from all three work together within the IRIS-HEP challenge format, the institute targets the development of solutions that can serve as a blueprint for the future.

The unifying IRIS-HEP vision for end-user physics analysis encompasses the following:
\begin{itemize}
    \item \textbf{Analyze $\mathcal{O}\left(1000~\textrm{TB}\right)$ of data within a few hours}: This scale captures the envisioned \ac{HL-LHC} physics analysis needs and multiple hours of turnaround time for a full pipeline run allow for efficient progress in a physics analysis.
    \item \textbf{Interactive analysis turnaround with a ``coffee break'' timescale}: The possibility of iterating on ideas on the timescale of a coffee break with a meaningful subset of data means that development can proceed rapidly and waiting time does not become a bottleneck.
    \item \textbf{Fully integrated \acp{AF}}: These facilities provide a convenient interfaces to access required services and computational resources.
    \item \textbf{User experience to empower big \& small teams}: A recent dedicated white paper~\cite{Ciangottini:2024vtl} describes desirable features from a user perspective, including the handling of software environments and collaboration with colleagues.
    \item \textbf{Easy access to state-of-the-art \ac{ML} techniques}: \ac{ML} techniques are widely used throughout physics analysis and new ideas have the potential to shape analysis workflows, so sufficient flexibility to incorporate novel developments is desirable.
    \item \textbf{Reproducibility, preservation, reuse}: Developing sustainable physics analyses will help maximize their long-term impact and legacy~\cite{Bailey:2022tdz}.
\end{itemize}

\section{The Analysis Grand Challenge}

\begin{figure}[ht]
    \centering
    \includegraphics[width=0.95\linewidth]{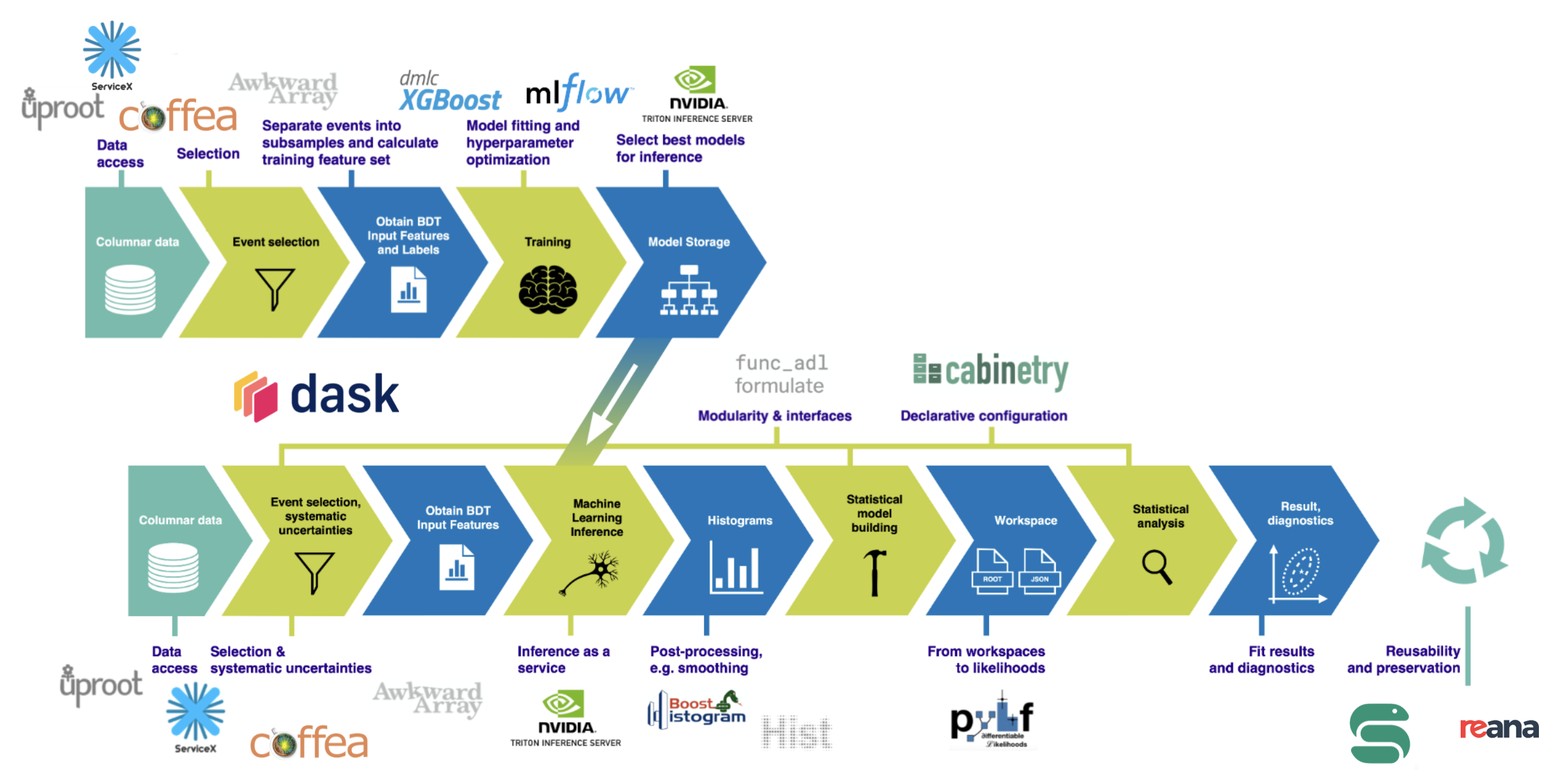}
    \caption{The IRIS-HEP reference implementation for the \ac{AGC} analysis task, figure adapted from~\cite{Held:2024gwj}.}
    \label{fig-agc-pipeline}
\end{figure}

The \ac{AGC} defines a physics analysis task starting from publicly accessible CMS Open Data~\cite{cms-open-data}.
It features the extraction and processing of data into histograms, \ac{ML} training and evaluation, as well as statistical inference.
The reference implementation of this task is provided by IRIS-HEP~\cite{agc_code} and heavily uses Python libraries from the Scikit-HEP~\cite{Rodrigues:2020syo} project.
This pipeline is depicted in Figure~\ref{fig-agc-pipeline}.

The \ac{AGC} project provides a useful environment for integration tests and a variety of implementations beyond the IRIS-HEP reference have been developed.
For IRIS-HEP, the implementation employs task graphs to express and execute data analysis operations.
This is done with Dask~\cite{dask}, a Python library providing both an interface to describe manipulations of data through task graphs, as well as ways to efficiently schedule the distributed execution of such graphs.
Deep integration of Dask with other \ac{HEP}-specific Python libraries, such as coffea~\cite{coffea}, has been achieved over time and Dask is additionally emerging as a common feature in many \acp{AF}.

The \ac{AGC} has been successful in its integration role but is limited in scale as the size of the input data is only around 2~TB.
This motivated the start of a new Grand Challenge project to specifically target scaling behavior.

\section{The 200 Gbps Challenge}

\begin{figure}[ht]
    \centering
    \includegraphics[width=0.85\linewidth]{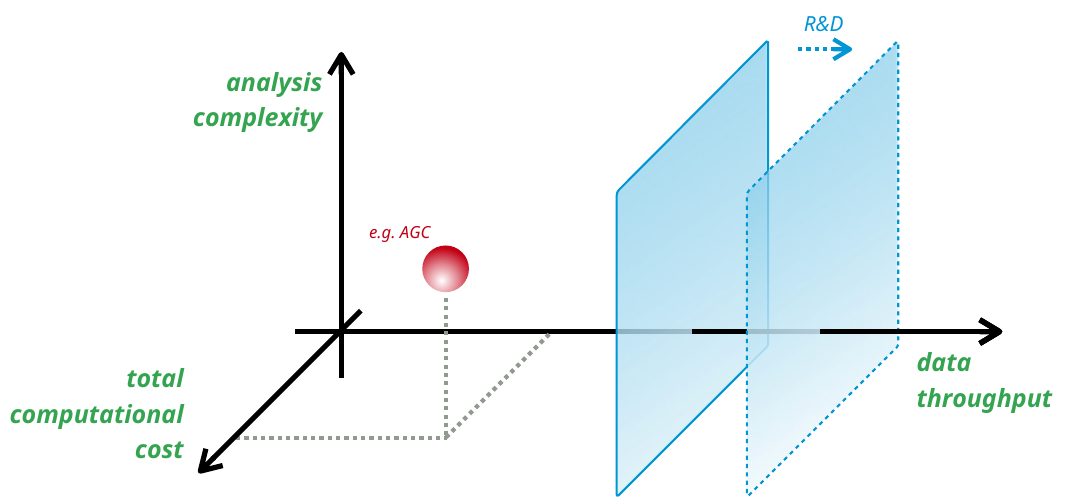}
    \caption{Factorizing computational aspects of physics analyses into independent challenges: the 200 Gbps project is focused on data throughput.}
    \label{fig-analysis-factorization}
\end{figure}

End user physics analysis pipelines are not centrally prescribed and there is limited agreement on how physics analyses will evolve towards the \ac{HL-LHC}.
It is therefore currently unclear what a ``representative'' \ac{HL-LHC} analysis might look like.
This poses a challenge for software and computing R\&D: in the absence of agreed-upon benchmarks, the evaluation of technical readiness is more difficult.
Figure~\ref{fig-analysis-factorization} proposes a way to factorize computational aspects of physics analysis pipelines into independent challenges.
One axis is pure data throughput, focused on how fast data can be read, decompressed and be made available for any subsequent physics analysis operations.
Another axis is the total computational cost of the pipeline: this can capture aspects such as the cost to train and evaluate \ac{ML} models, but also the processing of systematic variations for the evaluation of associated uncertainties in the physics result.
The third axis is the analysis complexity, which can relate to the number of different steps required in the workflow or the types of operations and external services that are required to be supported.
A given analysis example, such as the \ac{AGC} shown in red, can be located somewhere within this space.

A benchmark analysis example like the \ac{AGC} probes a specific point in this space.
Defining technical capabilities today, as well as moving forward towards the \ac{HL-LHC}, requires addressing what region of this space is technologically accessible.
The 200 Gbps Challenge project focuses purely on data throughput to evaluate what can be achieved with the approaches envisioned by IRIS-HEP and to subsequently push these boundaries further outwards with targeted development.

\subsection{Defining the challenge}

The 200 Gbps Challenge is named after its key goal: demonstrate data throughput in a physics analysis context at a rate of 200 Gbps, sustained over half an hour.
These numbers target physics analysis at scale to probe \ac{HL-LHC} needs.
A given physics analysis will frequently only read a subset of the content in the centrally provided input files.
Given a target 25\% of the file content needing to be read, a rate of 200 Gbps corresponds to processing a 180 TB dataset within half an hour.

The relevant data formats in the \ac{HL-LHC} context for the CMS and ATLAS experiments are NanoAOD~\cite{nanoaod} and PHYSLITE~\cite{physlite}.
Table~\ref{table-200-gbps-numbers} provides more context about what this challenge means with event sizes typical for these formats.
For the NanoAOD example, reaching 200 Gbps with 1000 cores means processing events with a rate of 50 kHz and a throughput of 25 MB/s per core.
For PHYSLITE, a slightly more complicated data structure, the use of 2000 cores would mean event rates of 5 kHz and throughput of 12.5 MB/s per core.
\begin{table}[ht]
    \centering
    \caption{Event numbers and rates for 200 Gbps data throughput sustained over half an hour for CMS NanoAOD and ATLAS PHYSLITE examples.}
    \label{table-200-gbps-numbers}
    \begin{tabular}{lrr}
        \toprule
        & CMS NanoAOD example & ATLAS PHYSLITE example \\
        \midrule
        size per event & 2 kB & 10 kB \\
        number of events & 90 billion & 18 billion \\
        target total event rate & 50 MHz & 10 MHz \\
        \bottomrule
    \end{tabular}
\end{table}

As the challenge targets data throughput in the context of subsequent physics analysis, it includes all the steps leading up to this: the initial read, network transfer, as well as decompression.
It captures the steps up until the point where arrays of data are available in memory, ready for any case-specific additional physics operations.

Two different setups were studied in the context of the 200 Gbps Challenge.
The CMS-targeted implementation~\cite{idap_cms_repo} uses Run-3 CMS NanoAOD data and was operated at an \ac{AF} located at the University of Nebraska--Lincoln.
The ATLAS version~\cite{idap_atlas_repo} uses Run-2 ATLAS PHYSLITE data and ran at the University of Chicago \ac{AF}.
While both implementations have a similar target of demonstrating sustained throughput of 200 Gbps, they do differ in various ways.
Most importantly, the input files vary in event size, object types, and compression used; the \acp{AF} differ in their setup and hardware.

The initial timeline for this project was tight: the goal was to present the status at the 2024 WLCG/HSF workshop, which took place 8 weeks after the creation of the challenge.
Despite this, the presentation~\cite{idap_wlcg_desy} at the workshop demonstrated that the goal was achieved and it was further improved upon subsequently.

\subsection{Key Analysis Facility elements}

\begin{figure}[ht]
    \centering
    \includegraphics[width=0.95\linewidth]{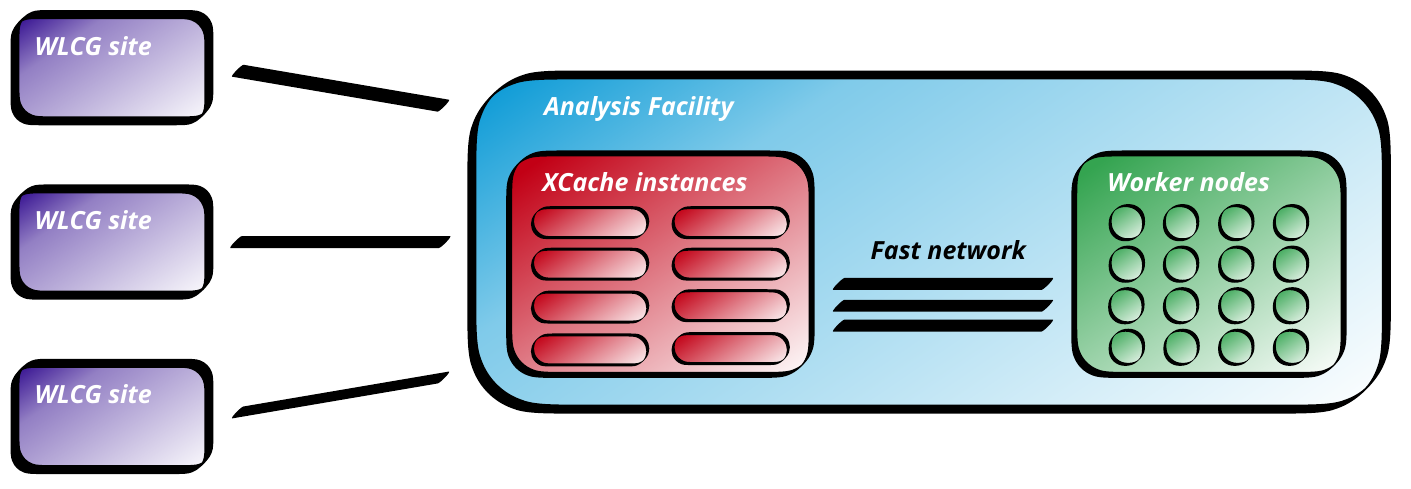}
    \caption{Key \ac{AF} components used to address the 200 Gbps Challenge.}
    \label{fig-af-schematic}
\end{figure}

Figure~\ref{fig-af-schematic} schematically depicts the data flow and relevant \ac{AF} components used in the challenge context.
The input data starts out distributed across the \ac{WLCG}.
Reading over the wide-area network puts limits on throughput and latency which are incompatible with the goals of this challenge.
XCache~\cite{CMS:2014cof,ATLAS:2018lpt} instances are deployed within the \acp{AF} to addresses this.
Remote data is cached during the first read and all subsequent reads happen out of the cache and are contained to the local network for fast and stable data access.
For the purpose of the following measurements all caches are assumed to be warm, with the caching having already taken place previously.
This corresponds to the typical situation during the development of physics analysis where physicists iterate on a design and will repeatedly run similar data processing pipelines.
The \acp{AF} also provide ways to distribute the workflow across available resources.
In the cases discussed here the individual worker nodes provide CPU resources for data handling and decompression.

\section{Measurements with the CMS setup at Nebraska}

Benchmarking with the CMS setup~\cite{idap_cms_repo} took place on the coffea-casa~\cite{coffea-casa} \ac{AF} located at the University of Nebraska--Lincoln.
Coffea-casa is built as a prototype for a \ac{HL-LHC} \ac{AF} with a hybrid setup employing both Kubernetes resources and resources via the local CMS Tier-2 and Tier-3 site.
Eight XCache instances were deployed, each with a 2x100 Gbps uplink, to scale to the intended throughput level.
Details about the setup can be found in a dedicated contribution to this conference~\cite{unl_idap}.

The measurements run a task graph built with Dask.
Nodes in this graph encapsulate the work to be done for a single file: reading the data with uproot~\cite{uproot} into awkward~\cite{awkward} arrays and returning metadata for benchmarking purposes.
Both Dask and TaskVine~\cite{10793142} were used to perform the distributed execution using resources provided by both HTCondor~\cite{10.5555/1064323.1064336} and Kubernetes.

\begin{figure}[ht]
    \centering
    \sidecaption
    \includegraphics[width=0.65\linewidth]{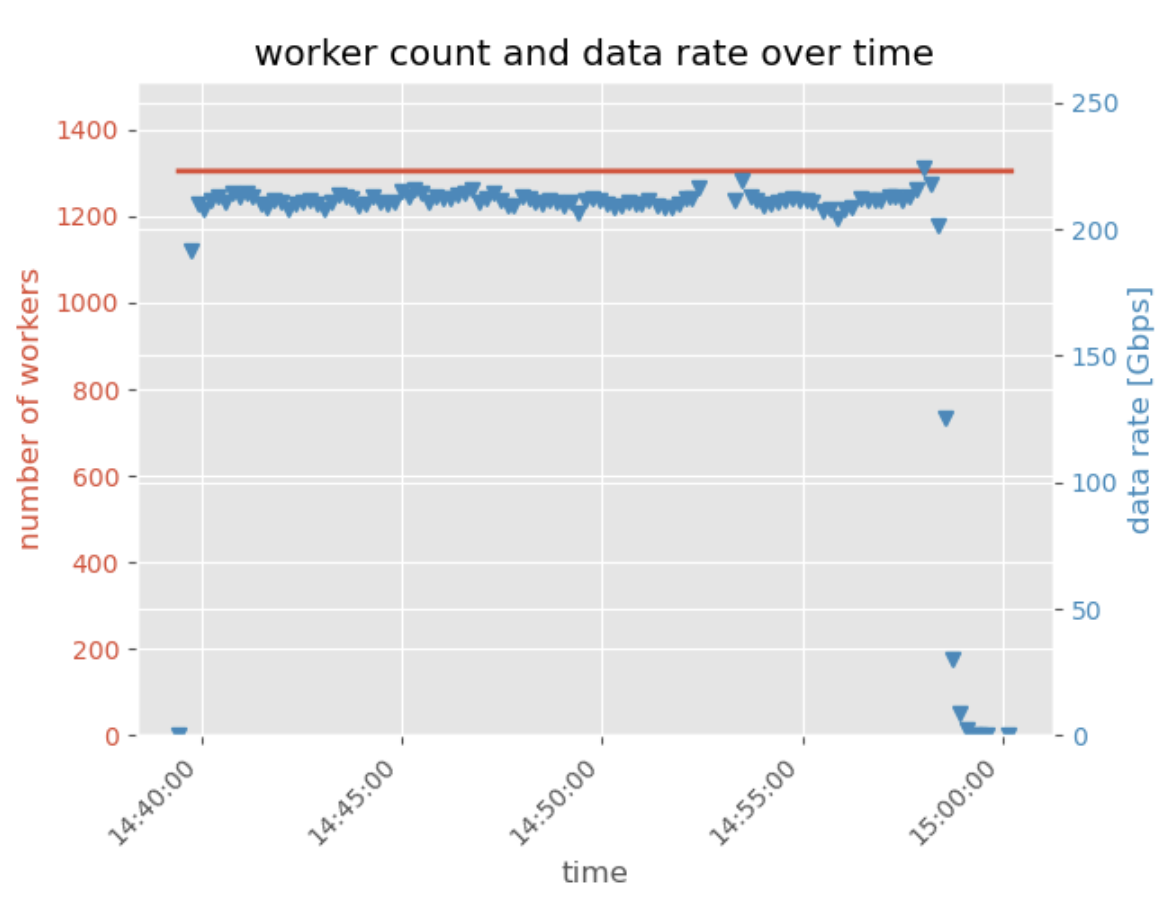}
    \caption{Data rate measurement at the coffea-casa \ac{AF} located at the University of Nebraska--Lincoln. A rate exceeding the target of 200 Gbps was sustained over 15 minutes in a distributed setup using around 1300 workers.}
    \label{fig-rate-nebraska}
\end{figure}

Figure~\ref{fig-rate-nebraska} shows a measurement example with Dask on HTCondor resources.
Using a constant allocation of around 1300 workers, a data rate exceeding 200 Gbps was sustained over a period of roughly 15 minutes.
This example processed 40 billion events distributed across 64k NanoAOD files for an event rate of 32 MHz.
A total of 30 TB of compressed data was read, corresponding to an uncompressed size of 71 TB.
This measurement was performed on NanoAOD inputs that were re-compressed from the default LZMA to instead use the Zstandard algorithm.
The change in compression algorithm resulted in an event rate increase per core of roughly a factor 2.5 and made reaching the 200 Gbps target possible.

\section{Measurements with the ATLAS setup at Chicago}

The ATLAS setup~\cite{idap_atlas_repo} used the University of Chicago \ac{AF}.
This \ac{AF} is used in production for ATLAS, is fully Kubernetes-based and was partially reconfigured to address the computational needs of the 200 Gbps Challenge.
A detailed description about the facility configuration is provided in a separate contribution to this conference~\cite{uchicago_idap}.
Eight XCache instances were deployed at this \ac{AF} as well, distributed across the network to maximize the available total bandwidth to the computing resources.

Two different data pipeline configurations were pursued: a Dask-distributed reading of data through an XCache with uproot and a setup employing ServiceX~\cite{servicex} as a data delivery service.

\subsection{Results with Dask and uproot}

\begin{figure}[ht]
    \centering
    \sidecaption
    \includegraphics[width=0.65\linewidth]{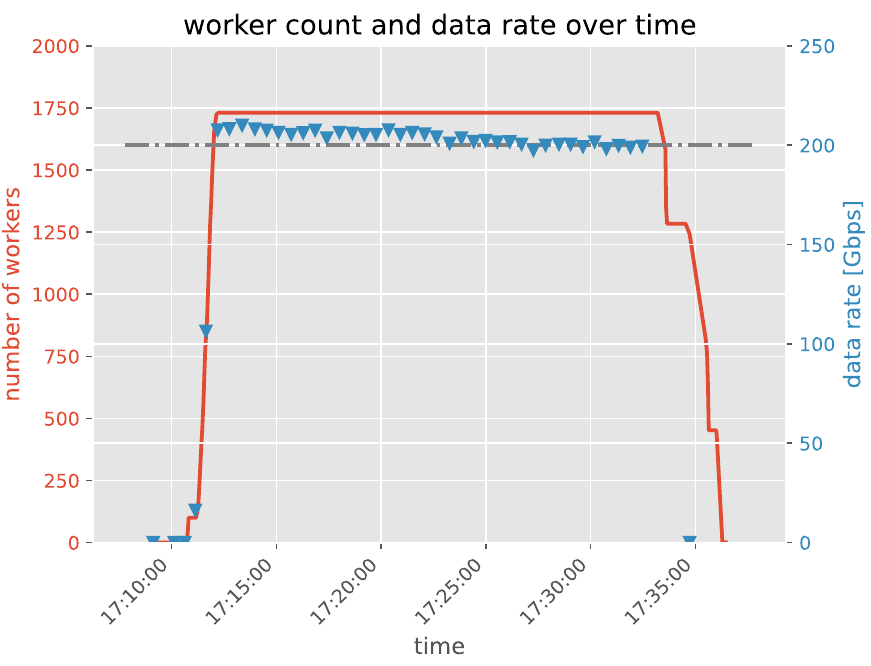}
    \caption{Data rate measurement at the University of Chicago \ac{AF}. A rate exceeding the target of 200 Gbps was sustained over 20 minutes in a distributed setup using up to around 1700 workers.}
    \label{fig-rate-chicago}
\end{figure}

Using Dask to distribute reading of data into awkward arrays with uproot, the target 200 Gbps data rate was sustained over a period of 20 minutes in the example shown in Figure~\ref{fig-rate-chicago}.
The worker allocation was dynamic and controlled by the Dask scheduler.
It quickly ramps up to a peak of around 1700 workers and scales down again towards the end of the execution as the number of remaining tasks decreases.
The example processed 32 billion events from 218k PHYSLITE files with a total size of 190 TB for an event rate of 15 MHz, with per-core rates of 5--20 kHz.
The uncompressed size of the data read was 32 TB, corresponding to 80 TB uncompressed.

\subsection{Results with ServiceX}

A second data processing pipeline that was tested for the challenge employed ServiceX as a data delivery service.
Data is read through the caches, processed by ServiceX, and then written out to object storage.
The resulting output can then be processed further in a next stage, for example with another Dask-distributed setup.
This multi-stage schema can be beneficial especially when the data size significantly decreases in the first processing step, for example due to event filtering being applied.
Repeated executions of the following processing step can then be particularly efficient if they only infrequently necessitate re-running the earlier step.

\begin{figure}[ht]
    \centering
    \sidecaption
    \includegraphics[width=0.6\linewidth]{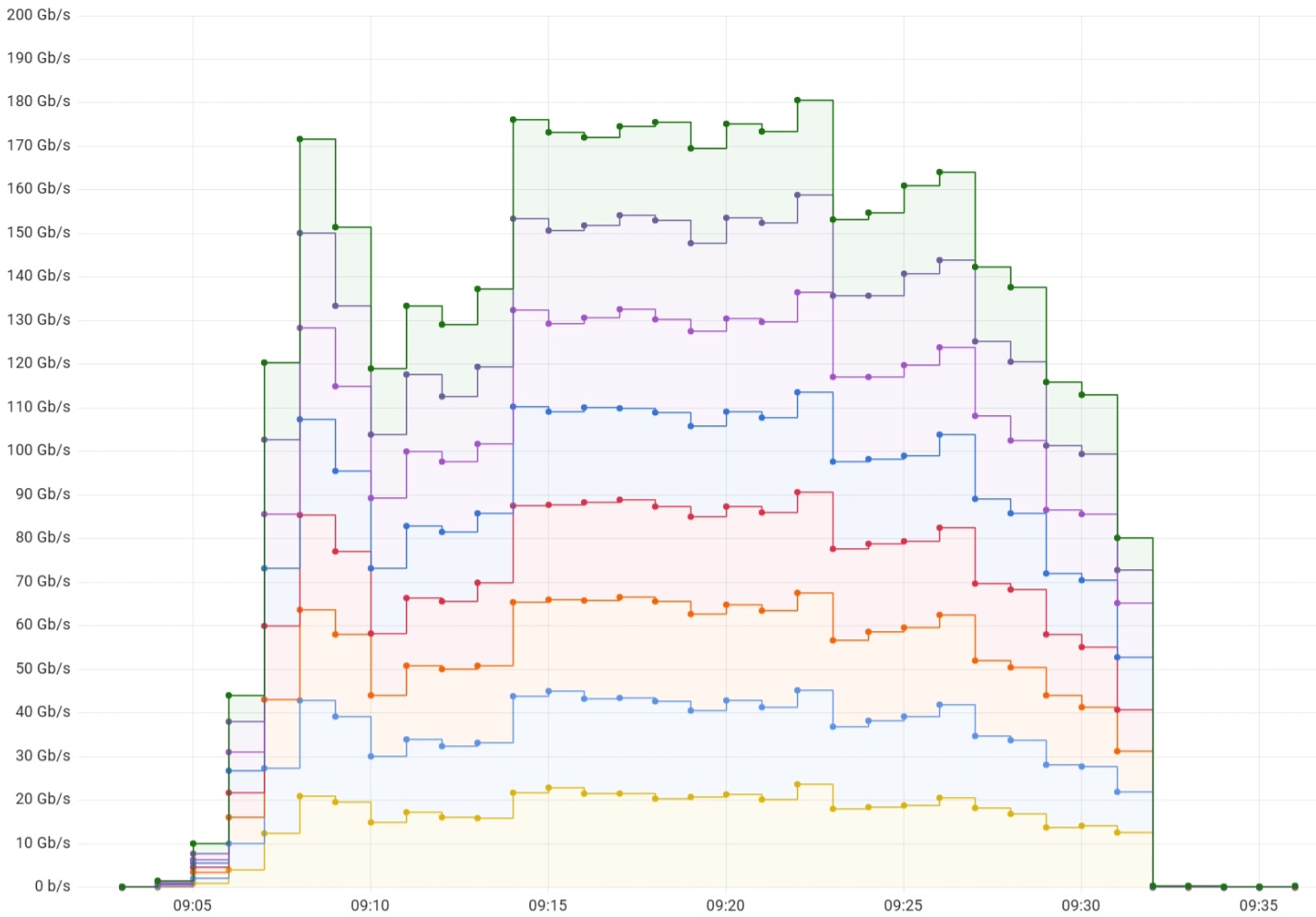}
    \caption{Data rate measurement at the University of Chicago \ac{AF}. Using ServiceX, the network monitoring shows rates up to 170 Gbps in this 25 minute example using up to 1000 pods.}
    \label{fig-rate-servicex}
\end{figure}

Figure~\ref{fig-rate-servicex} shows the data rates achieved in an example run, measured through network monitoring.
This example resulted in a rates of 170 Gbps over a 25 minute runtime.
It processed 19 billion events corresponding to 146 TB of data, using up to 1000 pods.
The processing time of subsequent analysis steps, which read data out of the object storage, strongly depends on the filtering performed during this ServiceX stage.

\section{Multi-user interactive analysis}

Future \acp{AF} are aiming to host many physicists who all would like to simultaneously scale to sufficiently many resources to enable interactive analysis turnaround.
It is therefore important to demonstrate that  good performance can also be achieved with multiple users running in parallel.
The two exercises reported here looked at different scenarios in this context at the \acp{AF} at the University of Nebraska--Lincoln and the University of Chicago.

\begin{figure}[ht]
    \centering
    \includegraphics[width=0.65\linewidth]{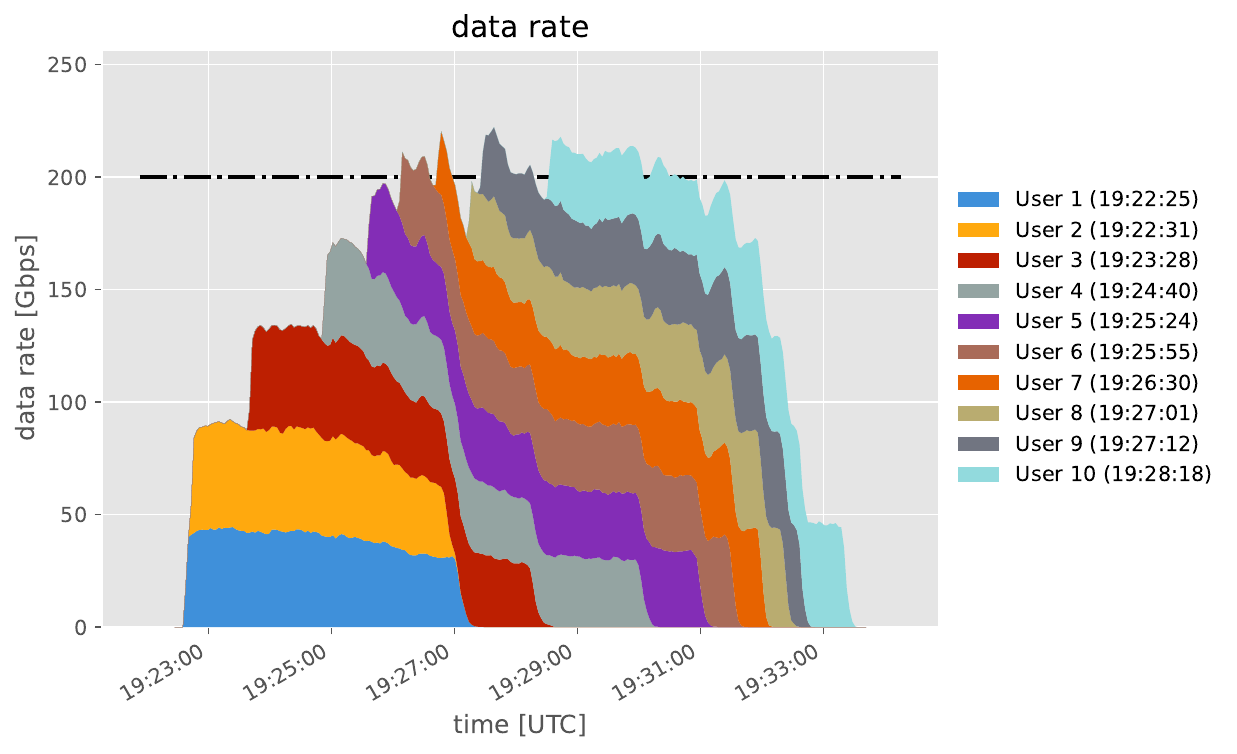}\\
    \includegraphics[width=0.65\linewidth]{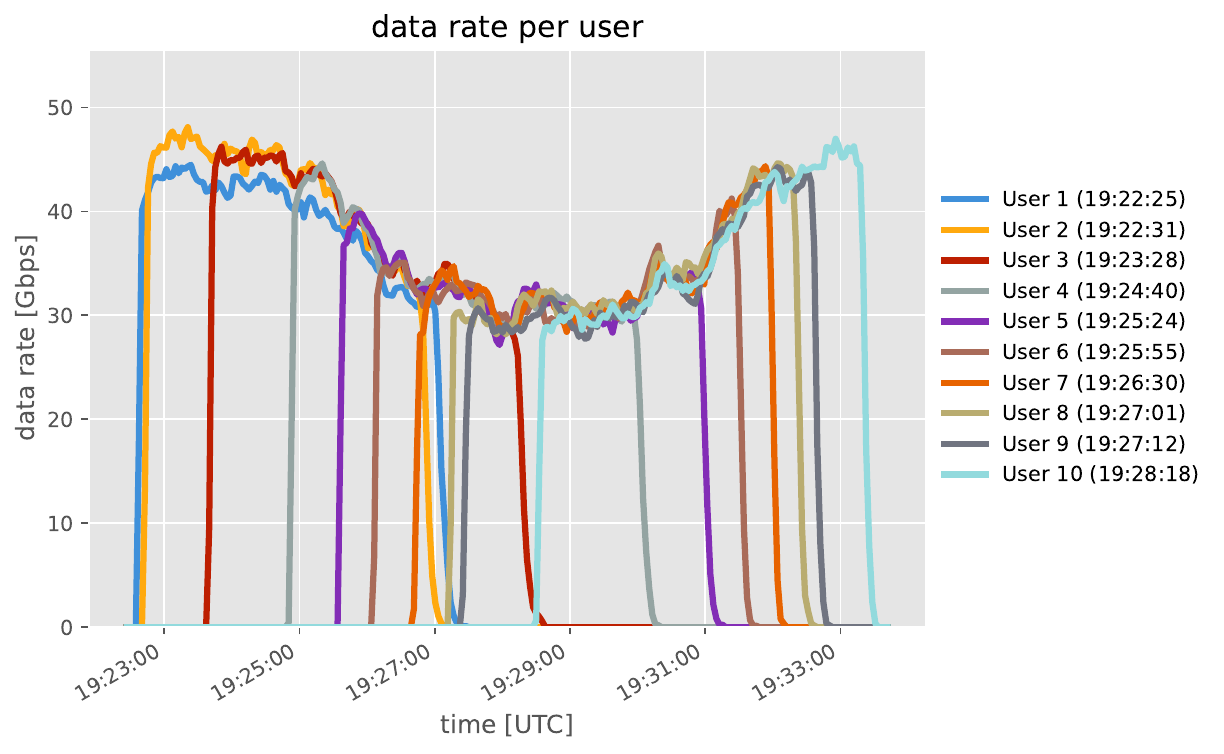}
    \caption{Multi-user test at the University of Chicago \ac{AF}. With ten users executing data processing pipelines, the available bandwidth was distributed between users and reached 200 Gbps (top). A saturation effect is visible in the per-user data rates (bottom).}
    \label{fig-multi-user-uchicago}
\end{figure}

The first exercise featured ten ATLAS users at the University of Chicago \ac{AF} launching data processing pipelines adapted from the 200 Gbps setup.
Launch times were randomized and each user was restricted to be assigned at most 200 cores, with Dask dynamically scaling the allocation.
The total data rate shown in Figure~\ref{fig-multi-user-uchicago} (top) reached 200 Gbps, distributed between multiple different users.
Figure~\ref{fig-multi-user-uchicago} (bottom) depicts the data rate per user as a function of time.
The rates dropped in the middle of the exercise as the network resources were saturated at this point in time.

\begin{figure}[ht]
    \centering
    \includegraphics[width=0.65\linewidth]{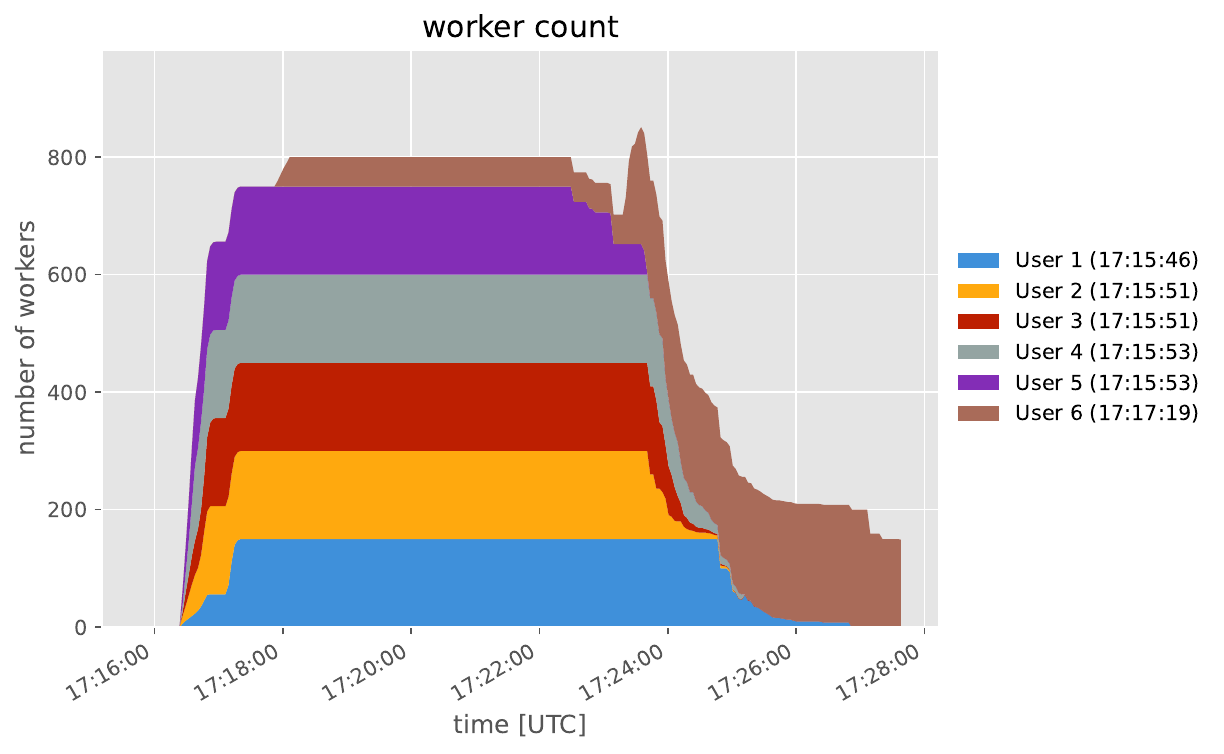}\\
    \includegraphics[width=0.65\linewidth]{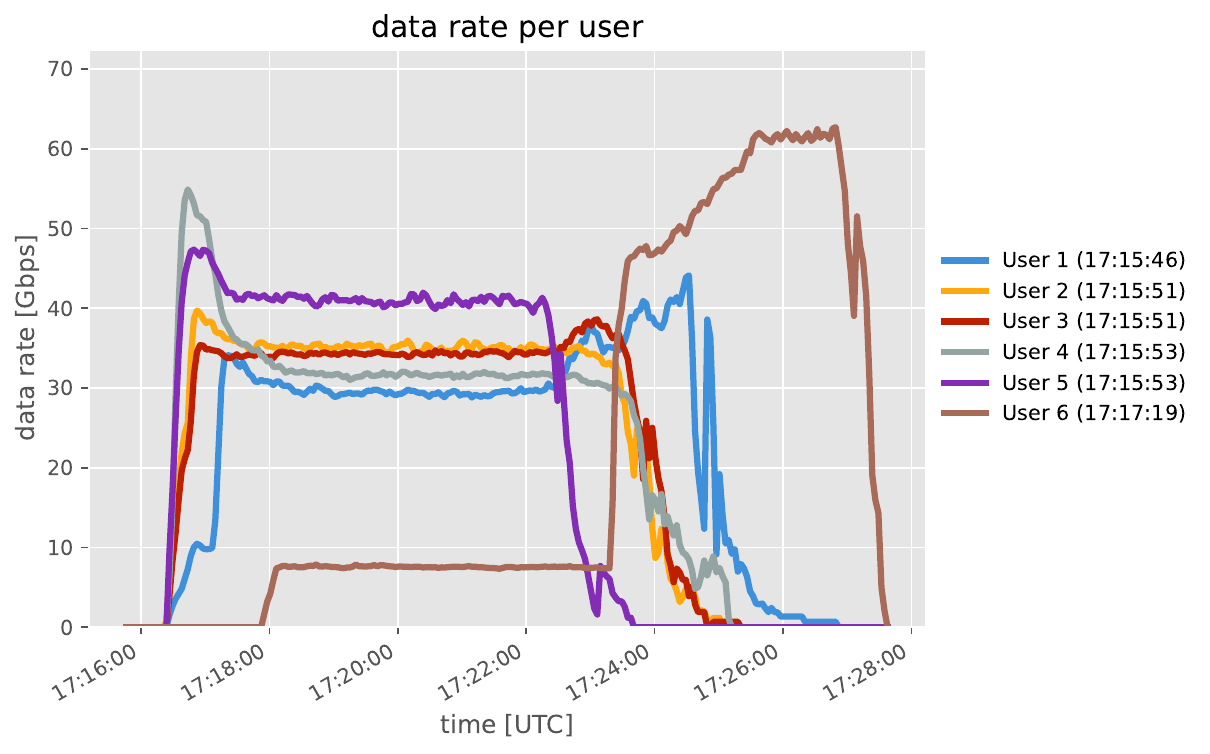}
    \caption{Multi-user test on the coffea-casa \ac{AF} at the University of Nebraska--Lincoln. When the resource demand for all users exceeded the available resources, users received more resources as other tasks finish up (top). The per-user data rates show the effect of resource redistribution and subsequent change in data rates (bottom).}
    \label{fig-multi-user-unl}
\end{figure}

The second exercise used the coffea-casa \ac{AF} at the University of Nebraska--Lincoln.
Figure~\ref{fig-multi-user-unl} (top) shows the worker allocation per participant.
Five CMS users launched a data processing pipeline at the same point in time and were allocated 150 workers each.
Dask dynamically scaled the allocation up to the 150 worker limit.
The sixth user received a more limited allocation due to the total availability of resources.
After the first tasks finished up, the sixth user was dynamically allocated more of the available resources through Dask.
The data rate for this user subsequently increased significantly as shown in Figure~\ref{fig-multi-user-unl} (bottom).

These measurements demonstrate how resources can be dynamically distributed across users.
Finding the right resource provisioning and fair-share patterns in the context of interactive analysis will likely require more work in this direction.

\section{Conclusion and next steps}

The 200 Gbps Challenge successfully demonstrated technology readiness and provided a checkpoint towards the \ac{HL-LHC}.
It served as a valuable mechanism to generate feedback and identify potential bottlenecks with large-scale data throughput analysis use cases.
The project relied on close collaboration between physicists, analysis software developers, and analysis facility experts in order to reach the 200 Gbps throughput target.

Looking ahead, IRIS-HEP intends to further study the space of physics analyses outlined in Figure~\ref{fig-analysis-factorization}.
Two directions are of particular interest: adding complexity to the existing 200 Gbps setup to capture more of the event-by-event processing of a physics analysis, as well as extending the throughput towards a 400 Gbps Challenge.
The future of \ac{HEP} end user physics analysis remains difficult to predict, but these targeted efforts aim to ensure readiness for a broad range of \ac{HL-LHC} use cases.

\begin{acknowledgement}
     This work was supported by the U.S. National Science Foundation (NSF) cooperative agreements OAC-1836650, OAC-2029176, OAC-2115148, PHY-2120747, PHY-2121686, and PHY-2323298.
\end{acknowledgement}

\bibliography{main}
\end{document}